\documentclass[final,3p,twocolumn,times]{elsarticle}

\usepackage{amssymb}

\usepackage{hyperref}

\usepackage{graphicx}

 \usepackage{threeparttable}
 \usepackage{multirow}

\usepackage[usenames]{color}
\usepackage{cancel}
\usepackage[normalem]{ulem}

\begin{document}

\begin{frontmatter}

\title{
Cycle frequency in standard Rock-Paper-Scissors games: \\ Evidence from experimental economics
}

\author{Bin Xu$^{1,2,3}$}

\author{Hai-Jun Zhou$^{3}$}

\author{Zhijian Wang$^{2,3}$\corref{cauthor}}

\cortext[cauthor]{Corresponding author. email: wangzj@zju.edu.cn}

\address{
$^{1}$  Public Administration College, Zhejiang Gongshang University,
Hangzhou, 310018, China
 \\
$^{2}$ Experimental Social Science Laboratory, Zhejiang University,
Hangzhou, 310058, China
 \\
$^{3}$ State Key Laboratory of Theoretical Physics, Institute of Theoretical Physics, Chinese Academy of Sciences, Beijing, 100190, China
}

\date{\today}

\begin{abstract}
  The Rock-Paper-Scissors (RPS) game is a widely used model system  in game theory.
  Evolutionary game theory predicts the existence of
  persistent cycles in the evolutionary trajectories of the RPS game, but
  experimental evidence  has remained to be rather weak.
  In  this work we performed laboratory experiments on the RPS game and analyzed
  the social-state evolutionary trajectories of twelve populations of $N=6$ players.
  We found strong evidence supporting the existence of persistent cycles.
  The mean cycling frequency was measured to be $0.029 \pm 0.009$ period per
  experimental round.   Our experimental observations can
  be quantitatively explained by a simple non-equilibrium model, namely the
  discrete-time logit dynamical process with a  noise parameter.
  Our work therefore favors the evolutionary game
  theory over the classical game theory for describing
  the dynamical behavior of the RPS game.
  \end{abstract}

\begin{keyword}
Rock-Paper-Scissors game;
cycle;
social state;
population dynamics;
evolutionary trajectory
\end{keyword}


\end{frontmatter}

\section{Introduction}

Evolutionary game theory (EGT) is becoming a general theoretical
framework to analysis strategies behaviors \cite{Maynard1982evolution,frey2010evolutionary,Friedman1998Rev}.
EGT is rooted in the classical game theory (CGT) \cite{VonNeumann1944} and
the theory of evolution \cite{darwin1859origin}.
Different from CGT, EGT predicts there could
exist persistent cycles
in the evolutionary trajectories in the strategy space \cite{Weibull1997,Sandholm2011,Nowak2012}.

As an example let us
consider the standard Rock-Paper-Scissors (RPS)
game. This is a prototype game in textbooks  \cite{Maynard1982evolution,VonNeumann1944,Weibull1997,nowak2006evolutionary,Sandholm2011}.
In this game, dynamics equations (e.g., the
standard replicator dynamics equations) in EGT
predict that the evolutionary trajectory will cycle around the
Nash equilibrium persistently.
However, the CGT predicts full random behavior:
the system is in  a mixed-strategy equilibrium
with each cyclic motion being balanced completely by its reverse cyclic motion.
According to the CGT theory,
cycles (also referred to as excess loops)
cannot be observed in the
evolutionary trajectories in the long run~\cite{Nowak2012,Friedman2012}.

Empirical examples of the RPS cycles are constantly being discovered
in nature, e.g., three morphs male lizard~\cite{Sin96} and others \cite{pryke2006red,Ker02}.
The environment for animal contests is decentralized, in which the encounter is pairwise, but population strategy shows cyclic behaviors \cite{Maynard1982evolution,Sin96}.
In general, human economic behaviors (e.g., exchanges) are also pairwise and not centralized \cite{hayek1945use}.
To test EGT in human game experiments~\cite{Falk2009},
the traditional setting is decentralized
(see review~\cite{Samuelson2002}), in which a subject in each round
competes with one
random-pairwise opponent
within a finite population
\cite{Friedman1996,Huyck1999,Friedman1998,HuyckSamuelson2001,Huyck2008,Binmore2001,XuWang2011ICCS}.
In such traditional setting experiment, whether the trajectories are persistent cycles instead of convergence to a Nash equilibrium remains an open question~\cite{Samuelson2002,benaim2009learning,Friedman2012}.
  Till now, no persistent cycle has been confirmed in the
  RPS experiment under such a traditional setting
  \cite{Nowak2012,Friedman2012}, and furthermore no dynamics observation
  has been  reported quantitatively.

In this paper we study the evolutionary trajectories of the
Rock-Paper-Scissors game from the perspective of non-equilibrium
statistical physics.
In non-equilibrium statistical physics studies,
formulating a physically meaningful measure of the
distance from equilibrium is an area of active research~\cite{sivak2012near}.
An equilibrium system satisfies the detailed balance condition, which
ensures the time reversal symmetry.
However, detailed balance is broken in a non-equilibrium system even in its
stationary state, therefore various dynamical patterns may show up in the
evolutionary trajectory.
Several non-equilibrium order parameters, such as entropy production~\cite{Evans2002experimental,Frey2010} and velocity~\cite{kumar2011symmetry},
have been constructed to characterize the distance from equilibrium.
In this work we carry out laboratory experiments on the RPS game, and
we detect the possible existence of persistent
cyclic flows using an angular frequency as the non-equilibrium
order parameter. A non-zero angular frequency serves as a
quantitative measure of the distance from equilibrium for the
evolutionary trajectories. We also compare our experimental observations
with the predictions of a simple non-equilibrium model, the
discrete-time logit dynamical process with a  noise parameter $\beta$.

Our experiments are the standard RPS games with the experimental setting
of discrete time, random pairwise matching
and local information. This setting has its reality in biology and economics \cite{Samuelson2002,hayek1945use,Huyck1999,Friedman1998,Nowak2012,Friedman1996,HuyckSamuelson2001,Huyck2008,Binmore2001,selten2008,Plott2008}.
We collect a total number of
twelve experimental trajectories from our experiments (each trajectory is the
result of $300$ rounds of the game) and then analyze these
trajectories.
Like other previous experiments
\cite{Friedman1996,Friedman1998,Huyck2008,Binmore2001,Milinski2003} and theories \cite{DynamicsinDiscreteTime2011,galla2009intrinsic},
the evolutionary trajectories are highly stochastic, but
 using our non-equilibrium order parameter
we are able to confirm that cycles exist and do not dissipate.
The mean frequency of cycles is about $0.029 \pm 0.009$ period per experimental round. This mean value is used to evaluate the noise parameter $\beta$
of the logit dynamics model, and a value of $\beta \approx 0.20$ is obtained.

This paper is organized as following. In the next section we introduce the
standard RPS game in the traditional setting and describe our data analysis protocol.
In section \ref{sec:results} we describe our main experimental results.
The experimental results are compared with the predictions of the
discrete-time logit dynamics model in section \ref{sec:model}.
We conclude this work in the last section.

\section{Experimental setup and data analysis}
\label{eq:method}

There are three different pure strategies in the Rock-Paper-Scissors game, namely
Rock ($R$), Paper ($R$) and Scissors ($S$). These three strategies form a
directed circle $R \rightarrow S \rightarrow P \rightarrow R$, namely
$R$ beats $S$,  $S$ beats $P$, and $P$ in turn beats $R$.
In our experiments we use the  simple
payoff matrix shown in
Table~\ref{RPSpaoffmatrix} to make the RPS game a constant-sum game:
In each play between two players, the winning player gets a payoff $2$
(i.e., two experimental points)
while the losing player gets a payoff $0$;
if there is a tie then each player gets an equal payoff $1$.

\begin{table}
\begin{center}
\begin{tabular}{cccc}
 ~~~& R & P & S\\
\end{tabular}\\
\begin{tabular}{c}
 R\\
 P\\
 S
\end{tabular}
\begin{tabular}{|c|c|c|}
\hline
1 & 0 & 2 \\
    \hline
2 & 1 & 0 \\
    \hline
0 & 2 & 1 \\
   \hline
\end{tabular}
\caption{\label{RPSpaoffmatrix}
Payoff matrix of the Rock-Paper-Scissors game.
The value of each matrix element is the payoff of the
row player's strategy given the strategy of the column player.}
\end{center}
\end{table}

\subsection{Experimental setting}

There were twelve independent and disjoint groups in our laboratory experiments.
Each group was formed by six players,
therefore the RPS game is a finite population game with population size $N=6$.
Each group played the RPS game $300$ rounds (we will explain the
motivation to use $300$ rounds later on). In each round of the play,
the six players of each group were first randomly assigned to three
disjoint sub-groups by a computer program, and then the two players of each
sub-group played the RPS game once.
All players made their own decisions simultaneously and anonymously.
After all the players
had submitted their choices,
each player then got the feedback information through her/his private computer screen.
The feedback information included her/his own strategy,
her/his opponent's strategy,
and her/his own payoff. No other information was provided to the players.
Each player also understood that her/his strategy in each round of the play is only
shown to her/his opponent of this round but not shown to the other players.

These $12$ experimental sessions were conducted during December 2010 in the
experimental social science laboratory of Zhejiang University. The $72$
experimental subjects (players) were recruited broadly from the student population
of the university.
They were sitting in an isolated seat with a computer
 during the games.
Both written and oral instructions were provided for each player before
the experiment. During the experiment, the players gained experimental points
in each round of the game according to the payoff matrix.
The experimental sessions lasted about $1.5$-$2$ hours.
The players got their earnings in cash privately after the experiment
according to the accumulated experimental points over the $300$ rounds.
The exchange rule is one experimental point equals $0.15$ Yuan RMB.
In addition, each player got $5$ Yuan RMB as show-up fee.
The average earning was about $50$ Yuan RMB.

\subsection{Data analysis}\label{sec:AataAnalysis}

There are three pure strategies in the RPS game, therefore
we use a vector $(x, y, z)$ to denote a generic \emph{social state} of
the population, with $x$, $y$ and $z$ being respectively
the fraction of players using strategy $R$, $P$ and $S$.
Suppose at the
$t$-th round of the game, $n_R(t)$ players used strategy $R$, $n_P(t)$
players used strategy $P$, and $n_S(t) = N-n_R(t)-n_P(t)$ players
used strategy $S$.
Then
$$
x \equiv \frac{n_R(t)}{N}\; , \quad
y \equiv \frac{n_P(t)}{N}\; , \quad
z\equiv \frac{n_S(t)}{N})\; .
$$
Obviously $x$, $y$, and $z$ should
satisfy $x\geq 0$, $y\geq 0$, $z\geq 0$, and
$x+y + z = 1$. The total number of different social states
for a population of size $N$ is simply $\frac{(N+1) (N+2)}{2}$.
In the studied case of $N=6$ this number is $28$.

\begin{figure}
  \begin{center}
    \includegraphics[width=1.0\linewidth]{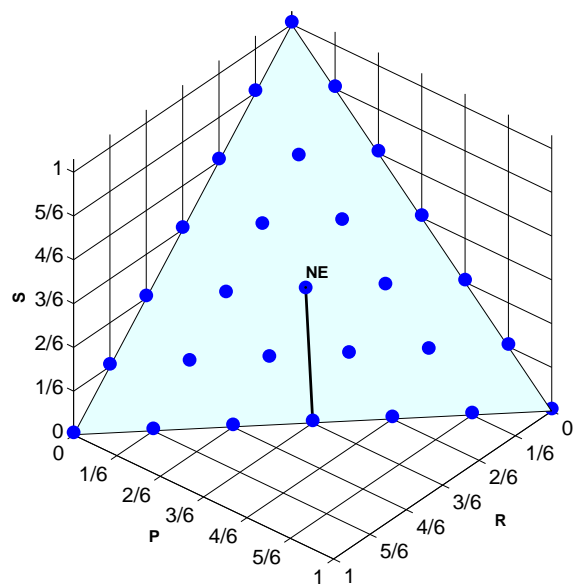}
  \end{center}
  \caption{
    Social state space and Poincar\'{e} section of the RPS game
    with $N=6$ players. Each social state is represented by a point $(x,y,z)$ on the
    plane characterized by $x+y+z=1$. All the $28$ social
    states are distributed within or on the boundary of
    an equilateral triangle. The social state
    $(\frac{1}{3}, \frac{1}{3}, \frac{1}{3})$
    is the Nash equilibrium (NE) point.
    To detect possible persistent flow  within the social state space,
    a line segment linking the Nash equilibrium point and the
    social state $(\frac{1}{2}, \frac{1}{2},0)$ is  drawn. This line is
    referred to as the Poincar\'{e} section.
  \label{fig:28state}
  }
\end{figure}

The social state $(x,y,z)$ of a population at a given time point
is a coarse-grained
description about the strategies used by the members of this population
\cite{XuWang2011ICCS,Nowak2012}.
The set of all the social states of a population
is referred to as the \emph{social state space} of the population. It can be
represented graphically by an equilateral triangle in a three-dimensional
Euclidian coordinate system, see Fig.~\ref{fig:28state}.
Each social state $(x, y, z)$ corresponds to a point in the interior or on the
boundary of this triangle. The central point
$(\frac{1}{3}, \frac{1}{3}, \frac{1}{3})$ of the triangle
is the Nash equilibrium (NE) point of the RPS game.

\begin{figure}
\begin{center}
    \includegraphics[width=1.0\linewidth]{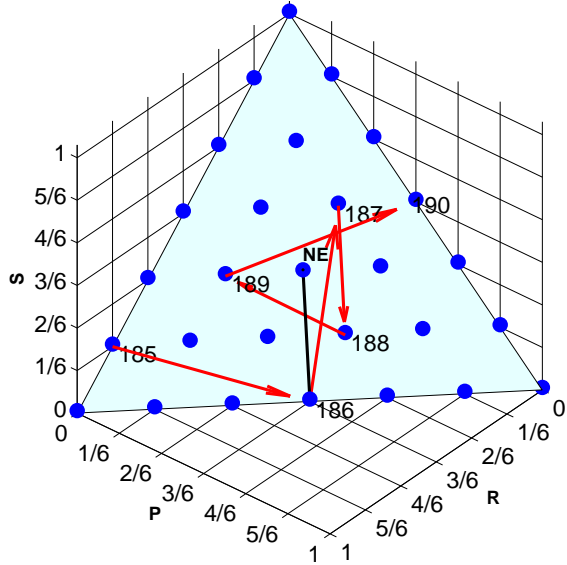}
\end{center}
\caption{
  A pictorial view of a short segment of an
  experimentally recorded evolutionary trajectory, starting from $t=185$ and
  ending at $t=190$. The counting numbers of the four social state transitions
  are, respectively, $C_{185} = 0.5$, $C_{186}= 0.5$, $C_{187}=0$,
  $C_{188}=-1$ and $C_{189}=0$. Therefore the accumulated counting number of this
  trajectory segment is $C_{185,190} = 0$.
  \label{fig:Transition}
}
\end{figure}

Generically speaking, the social state of the population
is different at different rounds $t$ of
the repeated RPS game. The social state $(x, y, z)$ as a function of the
discrete time $t$ forms an \emph{evolutionary trajectory} in the
social state space
\cite{XuWang2011ICCS,Nowak2012,Friedman2012,Binmore2001},
see Fig.~\ref{fig:Transition} for a simple illustration.
After an evolution trajectory of $T$ time steps has been collected,
we then perform
statistical analysis on it.
The first quantities of interest are the mean values of
$x$, $y$ and $z$, namely
\begin{equation}
\hspace*{-0.5cm}
\overline{x} \equiv \frac{1}{T} \sum\limits_{t=1}^{T} x(t)\; , \
\overline{y} \equiv \frac{1}{T} \sum\limits_{t=1}^{T} y(t)\; , \
\overline{z} \equiv \frac{1}{T} \sum\limits_{t=1}^{T} z(t)\; .
\label{eq:mean}
\end{equation}

To detect weak but persist directional motion in the social state space, we follow
Ref.~\cite{Friedman2012,Sven2011NplusNminus} and set a
line segment between the Nash equilibrium point and a point chosen at the boundary of
the triangle (see Fig.~\ref{fig:28state}). Such a line segment is referred to as
a Poincar\'e section.  Consider two consecutive social states
$ \vec{s}(t) \equiv \bigl(x(t), y(t), z(t)\bigr)$ and $
\vec{s}(t+1)\equiv \bigl(x(t+1), y(t+1), z(t+1)\bigr)$.
If either $\vec{s}(t)$ or $\vec{s}(t+1)$ is identical to the Nash equilibrium point, or
if the line segment from $\vec{s}(t)$ to $\vec{s}(t+1)$ does not cross the
Poincar\'e section, then the transition $\vec{s}(t) \rightarrow \vec{s}(t+1)$ is
assigned a counting number $C_t=0$. Otherwise,
(1) if the transition $\vec{s}(t)\rightarrow  \vec{s}(t+1)$
crosses the Poincar\'e section  from left to right (count-clockwise with
respect to the direction axis $(\frac{1}{\sqrt{3}}, \frac{1}{\sqrt{3}},
\frac{1}{\sqrt{3}})$ of the
social state plane),
then  $C_t=+1$;
(2) if this transition crosses the Poincar\'e section from right
to left (clockwise), then  $C_t=-1$;
(3) if $\vec{s}(t)$ is on the Poincar\'e section but is different from the
Nash equilibrium point, then $C_t=0.5$ ($C_t=-0.5$)
if  $\vec{s}(t+1)$ is to the right (left) of
the Poincar\'{e} section;
(4) if $\vec{s}(t+1)$ is on the Poincar\'e section but is different from
the Nash equilibrium point, then $C_t=0.5$ ($C_t=-0.5$)
if $\vec{s}(t)$ is to the left (right) of the Poincar\'e section.
We give some concrete examples of computing $C_t$ in
Fig.~\ref{fig:Transition}.

The accumulated counting number $C_{t_0, t_1}$ of the evolutionary
trajectory during the time interval $[t_0,t_1]$  is defined as
\begin{equation}
\label{Ce}
 C_{t_0,t_1} \equiv \sum\limits_{t=t_0}^{t_1-1} C_t \; .
\end{equation}
The accumulated counting number $C_{t_0, t_1}$ quantifies the
\emph{net} number of cycles around the Nash equilibrium point.
Such a quantity can help us to detect deterministic behaviors in a stochastic process
\cite{Sven2011NplusNminus}.
Starting from the initial time $t_0=1$, if
$C_{1, t}$ scales linearly with $t$ during the social-state evolution process, then
it indicates the existence of persistent cycles around the Nash equilibrium;
if $C_{1,t}$ as a curve of $t$ only fluctuates around $0$, then there is no
persistent cycles around the Nash equilibrium.
The mean frequency of cyclic motion in the time interval $[t_0, t_1]$ is
defined as
\begin{equation}
    \label{eq:expfrequency}
    f_{t_0, t_1} \equiv \frac{C_{t_0, t_1}}{t_1 - t_0} =
    \frac{1}{t_1 - t_0} \sum\limits_{t=t_0}^{t_1-1} C_t \; .
\end{equation}
Starting from the initial time $t_0=1$, we are interested in the value of
$f_{1, t}$ as $t$ becomes large.

\section{Experimental results}
\label{sec:results}

\begin{table}
\caption{\label{MeanSes12}
Statistics on the strategies. $\#R$ denotes the total number of times the
strategy $R$ being chosen by members of a given population ($\#P$ and $\#S$ have
similar meanings). $\overline{x}, \overline{y}, \overline{z}$ are defined by
Eq.~(\ref{eq:mean}).}
\begin{center}
\scriptsize
\begin{tabular}{|c|c|c|c|c|c|c|}
\hline
group & $\#R$ &	$\#P$ &	$\#S$ &	$\overline{x}$ & $\overline{y}$ & $\overline{z}$ \\
\hline
1 &	675&	601&	524&	0.375&	0.334&	0.291\\
2 &	632&	533&	635&	0.351&	0.296&	0.353\\
3 &	584&	591&	625&	0.324&	0.328&	0.347\\
4 &	688&	615&	497&	0.382&	0.342&	0.276\\
5 &	669&	568&	563&	0.372&	0.316&	0.313\\
6 &	642&	578&	580&	0.357&	0.321&	0.322\\
7 &	606&	583&	611&	0.337&	0.324&	0.339\\
8 &	625&	558&	617&	0.347&	0.31&	0.343\\
9 &	675&	581&	544&	0.375&	0.323&	0.302\\
10 &	604&	604&	592&	0.336&	0.336&	0.329\\
11 &	643&	567&	590&	0.357&	0.315&	0.328\\
12 &	659&	558&	583&	0.366&	0.31&	0.324\\
\hline
\end{tabular}
\end{center}
\end{table}

Table~\ref{MeanSes12} lists the total number of times the
three strategies
have been used in each of the $12$ evolutionary trajectories.
Among the twelve evolutionary trajectories of length $300$,
the total number of times the strategy $R$, $P$ and $S$
being used is, respectively, $7702$, $6937$ and $6961$.
The mean value of $x$, $y$ and $z$ as defined in Eq.~(\ref{eq:mean}), is
then $\overline{x} = 0.357 \pm 0.005$, $\overline{y} = 0.321 \pm 0.004$ and
$\overline{z} = 0.322 \pm 0.007$ (the standard deviation is estimated over
the $12$ evolution trajectories, see Table~\ref{MeanSes12}).
The observed mean point
$(\overline{x}, \overline{y}, \overline{z})$ is only slightly
different from the theoretical
Nash equilibrium point $(\frac{1}{3}, \frac{1}{3}, \frac{1}{3})$.

\begin{figure}
  \begin{center}
    \includegraphics[width=1.0\linewidth]{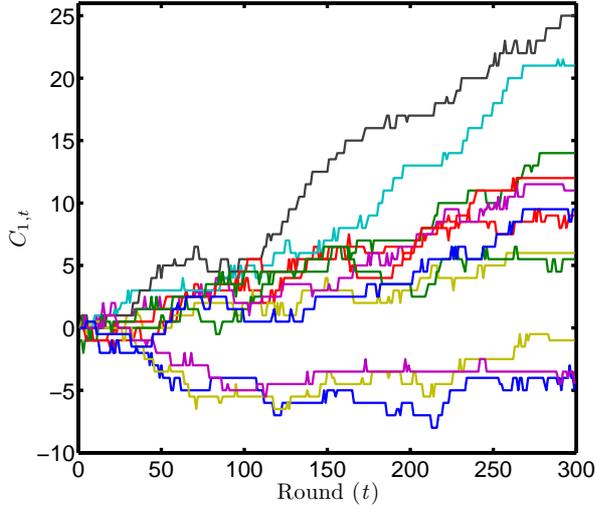}
  \end{center}
  \caption{
    Accumulated counting number $C_{1, t}$ as a function of evolution time $t$. Each of the twelve curves corresponds to one evolutionary trajectory involving six players.
   \label{fig:Trajectories}
  }
\end{figure}

The experimental trajectories are highly stochastic, similar to the
observations on other game processes
\cite{Friedman1996,Friedman1998,Huyck2008,selten2008,Binmore2001}.
However, if we
plot the evolution behavior of the
accumulated counting number $C_{1, t}$ with time $t$ in
Fig.~\ref{fig:Trajectories}, we find that $C_{1,t}$ increases with
$t$ in most of the data sets.
The value of $C_{1,300}$ for each of the $12$ experimental trajectories is
shown in the last column of Table~\ref{tab:Trajectories}. We obtain that the
mean value of $C_{1, 300}$ to be
$\overline{C}_{1, 300} =   8.54 \pm 2.66$.
Accordingly, the mean cycling frequency of these $12$ evolutionary
trajectories in $300$ steps is $\overline{f}_{1, 300} = 0.029 \pm 0.009$.
In other words, the empirical frequency of the cycles is $0.029 \pm 0.009$ period
per experimental round. The $95\%$ confidence interval of
this frequency  is $[0.009, 0.048]$.

Statistical analysis on the $12$ sampled values of $C_{1, 300}$ suggests that
the null hypothesis $H_1$ that $C_{1, 300}=0$ can be rejected
($p<0.01$, $t$-test).
Therefore we can say that cycles do exist in the RPS game
 in our experiments.
Statistical analysis also shows that $C_{1,300}>0$ ($p<0.01$),
i.e., the cycles are counter-clockwise around the Nash equilibrium point.
This result is consistent with the theoretical predictions of
some evolutionary dynamics models
\cite{Nowak2012,Maynard1982evolution,Sandholm2011}.

According to the last row in Table~\ref{tab:Trajectories}, to confirm the existence of
cycles using 12 samples,
 the trajectory length $t$ should be
at least $150$. This is because the
the null hypothesis ($C_{1, t}=0$) can be rejected ($p<0.05$)
 only when $t\geq 150$. That long
evolutionary trajectories are needed to confirm the existence of cycles can also
be understood from the empirical fact that the mean cycling frequency is very small.

To see the persistence of cycles,
setting null hypothesis as $C_{1,150} > C_{151,300}$ which means the cycles is disappearing along time. This hypothesis can be rejected by experimental
data ($p=0.06 < 0.10$).
Setting  $C_{1,100} > C_{201,300}$, this null hypothesis can be rejected
strongly ($p < 0.01$).
Concerning the question ``Do cycles dissipate when subjects must choose
simultaneously?" raised recently by the authors of Ref.~\cite{Friedman2012},
our experimental data therefore suggest that cycles do not dissipate.

\begin{table}
\begin{center}
\scriptsize
\begin{threeparttable}
  \caption{
    \label{tab:Trajectories}
    The accumulated counting number $C_{1,t}$ in the twelve evolutionary trajectories.
  }
  \begin{tabular}{|c|rrrrrr|}
   \hline
    Group	&$C_{1,50}$		&$C_{1,100}$	&$C_{1,150}$	 &$C_{1,200}$		 &$C_{1,250}$	 &$C_{1,300}$		\\
    \hline
	 1 & 0.50 & 3.00 & 5.50 & 5.50 & 8.50 & 9.00 \\
	 2 & 0.00 & 2.00 & 4.00 & 3.00 & 5.00 & 6.00 \\
	 3 & 1.00 & 1.50 & 4.50 & 7.00 & 10.00 & 14.00 \\
	 4 & 1.50 & 2.00 & 3.00 & 6.50 & 9.50 & 11.00 \\
	 5 & -3.00 & -4.00 & -5.00 & -6.00 & -4.00 & -5.00 \\
	 6 & 3.00 & 4.50 & 7.00 & 13.00 & 18.00 & 21.00 \\
	 7 & 4.00 & 5.50 & 12.50 & 17.00 & 21.00 & 25.00 \\
	 8 & 1.50 & 4.50 & 5.50 & 6.00 & 11.00 & 12.00 \\
	 9 & -3.50 & -5.50 & -3.50 & -4.50 & -2.50 & -1.00 \\
	 10 & 1.50 & 4.50 & 6.50 & 3.50 & 5.50 & 5.50 \\
	 11 & -2.00 & -5.00 & -3.50 & -3.50 & -3.50 & -4.50 \\
	 12 & -0.50 & 0.50 & 2.50 & 3.50 & 6.50 & 9.50 \\
\hline
    Mean & 0.33 & 1.13 & 3.25 & 4.25 & 7.08 & 8.54 \\
95\%L & -1.12 & -1.35 & 0.03 & -0.04 & 2.09 & 2.68 \\
     95\%U& 1.78 & 3.60 & 6.47 & 8.54 & 12.07 & 14.40 \\
$p$-value& 0.62 & 0.34 & 0.05 & 0.05 & 0.01 & 0.01 \\
\hline
  \end{tabular}
    \begin{flushleft}
        The last four rows are the statistical results of the $12$ experimental
        groups above. The row titled as $p$-value is $t$-test result by setting the null
        hypothesis $C_{1,t}=0$ for the $12$ samples.
        95\%U(L)  means the upper (lower) bound of $95\%$ confidence interval over
         the $12$ samples.
    \end{flushleft}
  \end{threeparttable}
  \end{center}
\end{table}

\section{Comparison with a simple model}
\label{sec:model}

To theoretically understand the experimental observations, we
now study a noisy best-response  process as a simple model for
the RPS game, namely the
discrete-time logit dynamics \cite{blume1993statistical}.
Multiple equilibria and limit cycles in the logit dynamics
has also been studied in a very recent paper by Hommes and Ochea
\cite{Hommes-Ochea-2012} in the continuous-time  limit.

Suppose the population of $N$ players is
in the social state $(x, y, z)$ after the $t$-th round of the game.
Let us denote by $u_i$ the mean payoff of the strategy $i\in \{R, P, S\}$ for
this social state. From the payoff matrix of Table \ref{RPSpaoffmatrix} we
can easily obtain that
\begin{equation}
u_R = x + 2 z  \; , \quad
u_P = y + 2 x  \; , \quad
u_S = z + 2 y \; .
\end{equation}
We assume that at the $(t+1)$-round of the game, each player of the population
will choose a strategy from $\{S, R, P\}$ independently of all the other players.
And we further assume that the time-dependent
probability $p_i$ for a player to choose strategy  $i$  is
\begin{equation}
\label{logitDyn}
p_i =
\frac{e^{ \beta u_i}}
{e^{\beta u_S} + e^{\beta u_R} + e^{\beta u_P}} \; ,
\quad\quad \forall i\in \{R, P, S\} \; .
\end{equation}
The parameter $\beta$ is referred to as
the ``inverse temperature" of the logit
dynamics, its value quantifies the rationality degree of
human agents in strategy interaction
\cite{blume1993statistical,Qlearning1992,mckelvey1995quantal,HuyckSamuelson2001,Wolpert2012,Hofbauer2003,Sandholm2011,updating2010,kianercy2012dynamics}.
In the limiting case of  $\beta=0$ each strategy will be
chosen with the uniform probability $\frac{1}{3}$.

For this simple Markovian process, the transition probability
$T^{(x, y, z)}_{(x^\prime, y^\prime, z^\prime)}$ from a social
state $(x, y, z)$ at time $t$ to another social state
$(x^\prime, y^\prime, z^\prime)$ at time $(t+1)$ is expressed as
(noticing that $z^\prime = 1-x^\prime-y^\prime$)
\begin{eqnarray}
 \hspace*{-0.4cm}
  T^{(x, y, z)}_{(x^\prime, y^\prime, z^\prime)}
& = & \frac{N!}{ (N x^\prime)! (N y^\prime)! (N z^\prime)!}
p_R^{N x^\prime} p_P^{N y^\prime} p_S^{N z^\prime} \nonumber \\
& = & \frac{N!}{ (N x^\prime)! (N y^\prime)! (N z^\prime)!}
\times \nonumber \\
& &  \frac{e^{N \beta[x x^\prime + y y^\prime + z z^\prime
+ 2 (  x y^\prime  +
y z^\prime  +  z x^\prime)]}}
{ (e^{ \beta (x + 2 z)} + e^{\beta (y + 2 x)}+ e^{\beta (z + 2 y )})^N}
\; .
\label{eq:Transition}
\end{eqnarray}
The steady-state probability $W_{(x, y, z)}^*$ that the
system is in the social state $(x, y, z)$ at $t=\infty$ can be
obtained by solving the following fixed-point equation
\begin{equation}
\label{eq:FixedPoint}
W_{(x, y, z)}^* =
\sum\limits_{(x^\prime, y^\prime, z^\prime)}
T^{(x^\prime, y^\prime, z^\prime)}_{(x, y, z)}
W_{(x^\prime, y^\prime, z^\prime)}^* \; .
\end{equation}
Because the transition probability from any social state
$(x, y, z)$ to any another social state $(x^\prime, y^\prime, z^\prime)$ is
positive, Eq.~(\ref{eq:FixedPoint}) has a unique solution with
the normalization property $\sum_{(x, y, z)} W_{(x, y, z)}^* = 1$
\cite{Kemeny-Snell-1983}. It is not difficult to prove  that the steady-state
probability distribution has the following rotational symmetry
\begin{equation}
\label{eq:rotation}
 W_{x, y, z}^* =   W_{y, z, x}^* = W_{z, x, y}^* \; .
\end{equation}
This rotational symmetry
 ensures that
\begin{equation}
\hspace*{-0.6cm}
\sum\limits_{(x, y, z)} x W_{(x, y, z)}^*
=
\sum\limits_{(x, y, z)} y W_{(x, y, z)}^*
=
\sum\limits_{(x, y, z)} z W_{(x, y, z)}^*
= \frac{1}{3} \; ,
\end{equation}
namely the logit dynamics will reach the Nash equilibrium $(\frac{1}{3},
\frac{1}{3}, \frac{1}{3})$ at $t\rightarrow \infty$.

It can be checked numerically and analytically that, for any $\beta > 0$,
the detailed balance condition is violated at the steady-state of the logit dynamics.
For two different social states $(x,y,z)$ and $(x^\prime, y^\prime, z^\prime)$, in
general we will find that
$$
T^{(x, y, z)}_{(x^\prime, y^\prime, z^\prime)}
W_{(x, y, z)}^*
\neq
T^{(x^\prime, y^\prime, z^\prime)}_{(x, y, z)}
W_{(x^\prime, y^\prime, z^\prime)}^* \; .
$$
Because of the violation of detailed balance, directional flows may
persist in the system even at $t\rightarrow \infty$.

\begin{figure}
    \begin{center}
       \includegraphics[width=1.0\linewidth]{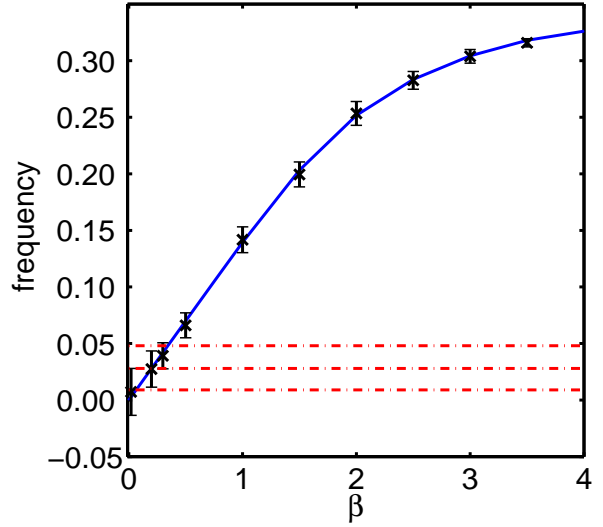}
    \end{center}
    \caption{
        \label{fig:angleN6}
        The steady-state mean cycling frequency $f^*$ of the discrete-time
        logit dynamical process with population size $N=6$. The solid line
        is theoretical result obtained with Eq.~(\ref{def:freq-theta});
        the cross ($\times$) symbols with error bars are obtained by averaging over
        many simulated evolutionary trajectories of length $300$.
        The mean experimental frequency of $\overline{f}_{1,300} \approx 0.029$ and
        its $95\%$ confidence upper and lower bound are
        marked by the dashed lines.
       }
  \end{figure}

We are especially interested in the directional flow around
the Nash equilibrium point.
Consider two social states $(x, y, z)$ and
$(x^\prime, y^\prime, z^\prime)$ on the
evolutionary trajectory at
two consecutive time points $t$ and $t+1$.
If either $(x, y, z)$ or $(x^\prime, y^\prime, z^\prime)$ is identical
to the Nash equilibrium point, the transition $(x, y, z)\rightarrow (x^\prime,
y^\prime, z^\prime)$ is \emph{not} a rotational motion
around the Nash equilibrium,
and we set the corresponding rotational
angle $\theta^{(x,y,z)}_{(x^\prime, y^\prime, z^\prime)}$ to be zero.
The Nash equilibrium point may be sitting on the rectilinear
line that passing through the social states $(x, y, z)$ and $(x^\prime, x^\prime, z^\prime)$. If this is the case,
the transition $(x,y,z)\rightarrow (x^\prime, y^\prime, z^\prime)$
is also \emph{not} a rotational motion around the Nash equilibrium, and
its rotational angle $\theta^{(x,y,z)}_{(x^\prime, y^\prime, z^\prime)}$
is again set to be zero. In all the remaining cases, the social states $(x, y, z)$, $(x^\prime, y^\prime, z^\prime)$ and the Nash equilibrium
point form a triangle in the
social state plane of Fig.~\ref{fig:28state}. The magnitude of the
rotational angle $\theta^{(x,y,z)}_{(x^\prime, y^\prime, z^\prime)}$ is
just the angle of this triangle at vertex point $(\frac{1}{3}, \frac{1}{3},
 \frac{1}{3})$, it must be
less than $\pi$. The rotational angle $\theta^{(x,y,z)}_{(x^\prime, y^\prime, z^\prime)}$ is defined as positive if the rotation from $(x, y, z)$ to $(x^\prime,
y^\prime, z^\prime)$ with respect to the Nash equilibrium point is counter-clockwise,
otherwise it is defined as negative.

At the steady-state of the discrete-time logit dynamics, the mean
frequency  $f^*$ that the evolution trajectory rotates around the Nash
equilibrium point can then be computed by the following formula
\begin{equation}
\label{def:freq-theta}
f^* \equiv
\frac{1}{2 \pi}
\sum\limits_{(x, y, z)} W_{(x, y, z)}^* \sum\limits_{(x^\prime, y^\prime, z^\prime)}
T^{(x, y, z)}_{(x^\prime, y^\prime, z^\prime)}
\theta^{(x,y,z)}_{(x^\prime, y^\prime, z^\prime)} \; .
\end{equation}
\begin{figure}
  \begin{center}
       \includegraphics[width=1.0\linewidth]{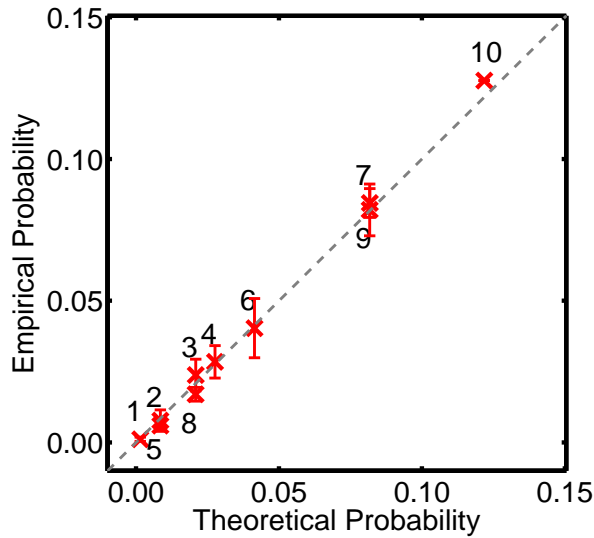}
  \end{center}
  \caption{
    \label{fig:Wvalues}
    Probability of staying in the social state $(x,y,z)$ for a population
    of size $N=6$.
    The horizontal axis is the predicted probability by the
    discrete-time logit dynamics model with inverse temperature
    $\beta = 0.20$, while the vertical axis is the
    empirical probability measured from the $12$ experimental
    trajectories.
    Because of the rotational symmetry (\ref{eq:rotation}),
    the $28$ social states
    can be coarse-grained into ten groups:
    (1), $\{(0,0,1), (0,1,0), (1,0,0)\}$;
    (2), $\{(0,\frac{1}{6},\frac{5}{6}),(\frac{1}{6}, \frac{5}{6},0),(\frac{5}{6},0, \frac{1}{6})\}$;
    (3), $\{(0,\frac{1}{3},\frac{2}{3}), (\frac{1}{3},\frac{2}{3},0),
    \frac{2}{3},0,\frac{1}{3})\}$;
    (4), $\{(0,\frac{1}{2},\frac{1}{2}),(\frac{1}{2},\frac{1}{2},0),
    (\frac{1}{2},0,\frac{1}{2})\}$;
    (5), $\{(\frac{1}{6},0,\frac{5}{6}), (\frac{5}{6}, \frac{1}{6}, 0),
    (0,\frac{5}{6}, \frac{1}{6})\}$;
    (6), $\{(\frac{1}{6},\frac{1}{6},\frac{2}{3}), (\frac{1}{6}, \frac{2}{3},
    \frac{1}{6}), (\frac{2}{3},\frac{1}{6}, \frac{1}{6})\}$;
    (7), $\{(\frac{1}{6},\frac{1}{3},\frac{1}{2}), (\frac{1}{3}, \frac{1}{2},
    \frac{1}{6}), (\frac{1}{2}, \frac{1}{6}, \frac{1}{3})\}$;
    (8), $\{(\frac{1}{3},0,\frac{2}{3}), (\frac{2}{3}, \frac{1}{3},
    0), (0, \frac{2}{3}, \frac{1}{3})\}$;
    (9), $\{(\frac{1}{3},\frac{1}{6},\frac{1}{2}), (\frac{1}{2}, \frac{1}{3},\frac{1}{6}
    ), (\frac{1}{6},\frac{1}{2}, \frac{1}{3})\}$;
    (10), $\{(\frac{1}{3}, \frac{1}{3},  \frac{1}{3})\}$.
    All the social states of a given group have the same
    stationary probability (the same
    horizontal-axis value)
    according to the theoretical model, but their measured probabilities
    might be different (the mean vertical-axis value and the standard error).
    }
\end{figure}

For the population size $N=6$, we show in Fig.~\ref{fig:angleN6} the steady-state
mean frequency $f^*$ as a function of the inverse temperature $\beta$.
To check the correctness of the theoretical calculations,
we also perform computer simulations based on
the discrete-time logit dynamics model to generate a set of simulated
evolutionary trajectories of length $300$. The mean cycling frequencies of these simulated evolutionary trajectories
are also shown in Fig.~\ref{fig:angleN6}.
The agreement between analytical
calculations and computer simulation results are very good.
We find that $f^*$ increases almost linearly with the inverse temperature
$\beta$ when $\beta<1.5$.
Comparing the theoretical results with the
mean frequency value of $\overline{f}_{1, 300}=0.029$, we infer the
inverse parameter should be set to $\beta = 0.20$.

At $\beta = 0.20$, we also perform computer simulations based on the
discrete-time logic dynamics model to generate a set of independent
evolution trajectories of length $T=300$. We then perform the same
analysis on these trajectories and find that the direction of the cycles is counter-clockwise and the mean cycling frequency is
$\overline{f} \approx 0.029$, consistent with the experimental result.

At $\beta=0.20$, the steady-state probability $W_{(x, y, z)}^*$ of
visiting each social state $(x, y, z)$ as predicted by the
logit dynamics is compared with the empirically observed probability
of visiting $(x, y, z)$, see Fig.~\ref{fig:Wvalues}.
The agreement between theory and experiment is again very good.

Although the discrete-time noisy-response logit dynamic model can
describe our experimental observations excellently, we should point out
an important difference between the model assumption and the experimental
setting. In our experiments, after each round of the game, each player only
knows the strategy of her/his opponent but not the social state of the whole
population. However in the logit dynamics model, we assume that each
player choose a strategy based on the knowledge of the current
social state of the population, see Eq.~(\ref{logitDyn}). In this sense,
the logit dynamics model is still a phenomenological  model. It is of
interest to quantitatively describe the RPS evolutionary dynamics by a
more microscopic model. We hope to return to this issue in a future study.

\section{Conclusion and discussions}

As a brief summary, in this work we studied the Rock-Paper-Scissors game
both experimentally and analytically. Our experimental data gave strong
evidence that counter-clockwise
cycles around the Nash equilibrium point exist in the social-state evolutionary
trajectory of a finte population.
We demonstrated that our experimental observations can be quantitatively
understood by a simple theoretical model of noisy-response logit dynamics.

RPS game experiments on EGT were also reported quite recently by Cason
 and co-authors and by Hoffman and co-authors \cite{Friedman2012,Nowak2012}.
The backgrounds and cutting edges of the experiment research are
well documented in these two references \cite{Friedman2012,Nowak2012}.
Compared with the decentralized setting of our present work, the
experimental environments of the RPS game in these two recent works \cite{Friedman2012,Nowak2012}
are all centralized: Instead of pairwise meetings,
 in all of the experiments reported in \cite{Friedman2012,Nowak2012},
  each subject competes against the choices of all other subjects.
However,  the decentralized setting (especially the random matching pairwise setting) is more closer to the natural environments
in biology and economics (e.g.,~\cite{Samuelson2002,hayek1945use}).
For example,
the encounters of male lizards are pairwise meetings~\cite{Sin96}.
For decentralized population RPS games, according to our knowledge,
the existence of persistent cyclic motions was not
confirmed by any previous laboratory experiments.

Going back to traditional
(decentralized)
setting experiments of the simplest RPS game,
the present work added strong evidence in favor of the existence of
persistent cycles.  As a fundamental observation on cycle, the
mean frequency of cycles was quantitatively meassured.
There are tens of dynamics models which have been  build to interpret cyclic behavior in RPS game, however there are rare quantitative observations
from real experiments. Quantitative measurements from experiments is important, without which to evaluate a dynamics equation precisely is almost impossible (or plausible).
As demonstrated, our experimental observations can be quantitatively
understood by a simple theoretical model of noisy-response logit dynamics.

We wish to emphasize two major points of our experimental approach.
First, by recording sufficiently long evolutionary trajectories,
we were able to detect weak deterministic motion in a highly stochastic process.
We noticed that cycles can only be confirmed ($p<0.05$) when
the
trajectories are longer than $150$ rounds in twelve samples.
Second,
 we focus on time asymmetry of social state transitions.
Importance of time asymmetry has been well emphasised in non-equilibrium
statistical physics~\cite{Sven2011NplusNminus,sivak2012near,evans1993probability}.
The frequency is observed from the loops out of detailed balance.

\section*{Acknowledgements}

We thank Ken Binmore for helpful discussion and Zunfeng Wang for technical assistance.
The  work of B.X. and Z.W. was supported by a grant
from the 985 Project
 at Zhejiang University and by SKLTP of
ITP-CAS
 (No.~Y3KF261CJ1). The work of H.J.Z. was
 supported by the Knowledge Innovation Program of Chinese
Academy of Sciences (No.~KJCX2-EW-J02) and the National Science
Foundation of China (grant No.~11121403 and 11225526).


\end{document}